\documentclass[aps,prl,twocolumn,showpacs,superscriptaddress,groupedaddress]{revtex4}  
\usepackage{graphicx,comment,float,braket,textcomp,amsmath}  
\usepackage{dcolumn}   
\usepackage{bm}        
\usepackage{amssymb}   

\begin{document}

\title{Vibrational Coupling Modifies Spectral Diffusion in Core-Shell Colloidal Quantum Dots}
\author{A. Liu} \affiliation{Physics Department, University of Michigan, Ann Arbor, Michigan 48109, USA}

\author{D. B. Almeida} \affiliation{Physics Department, University of Michigan, Ann Arbor, Michigan 48109, USA}

\author{W. K. Bae} \affiliation{SKKU Advanced Institute of Nano Technology, Sungkyunkwan University, Gyeonggi, Republic of Korea}

\author{L. A. Padilha} \affiliation{Instituto de Fisica ``Gleb Wataghin", Universidade Estadual de Campinas, 13083-970 Campinas, Sao Paulo, Brazil}

\author{S. T. Cundiff}
\email{cundiff@umich.edu}
\affiliation{Physics Department, University of Michigan, Ann Arbor, Michigan 48109, USA}

\date{\today}

\begin{abstract}
We perform two-dimensional coherent spectroscopy on CdSe/CdZnS core/shell colloidal quantum dots at cryogenic temperatures. In the two-dimensonal spectra, sidebands due to electronic coupling with CdSe lattice LO-phonon modes are observed that have evolutions deviating from the exponential dephasing expected from Markovian spectral diffusion. Comparison to simulations provides further evidence that the LO-phonon mode coupling itself significantly modifies the exciton lineshapes and results in the underlying spectral diffusion on ultrafast timescales.
\end{abstract}

\maketitle

Colloidal quantum dots (CQDs), which are semiconductor nanocrystals dispersed in solution, are the continued focus of intense study due to their applications in numerous areas including biological tagging \cite{Medintzetal2005}, next-generation display technologies \cite{Jangetal2010}, and high-efficiency photovoltaics \cite{Lanetal2014}. The comprehensive understanding of the exciton dynamics required for applications of CQDs is hindered by complexities resulting from electronic states coupling to external degrees of freedom, such as vibrational modes or different charge configurations.

Fluctuations in the exciton resonance energy due to interactions with the local environment, called spectral diffusion, are still not well understood. Studying spectral diffusion is challenging, since most spectroscopic studies on CQD ensembles have utilized one-dimensional techniques for which much of the desired information is obscured by the line broadening due to dot size dispersion. Much of the current knowledge about spectral diffusion is thus from experiments on single CQDs \cite{Ferneeetal2014-1}, which circumvent ensemble size inhomogeneity at the expense of low signal to noise ratio and dot-to-dot structural variations (oblateness, shell thickness/composition etc.). Current single-dot spectroscopy techniques are further limited in their time-resolution, with recent photon-correlation Fourier spectroscopy measurements \cite{Coolenetal2008,Beyleretal2013} probing spectral diffusion dynamics down to the $\mu$s timescale.

Multi-dimensional coherent spectroscopy (MDCS) \cite{CundiffMukamel2013} is a technique able to unfold the optical response of an inhomogeneous ensemble of emitters with femtosecond time resolution. MDCS correlates absorption, intraband (Raman) \cite{FerrioSteel1998}, and emission spectra. By simultaneously resolving the response of all constituent frequency groups within the excitation bandwidth, the homogeneous response of inhomogeneously broadened systems may be efficiently studied by MDCS. As examples, MDCS has recently been applied towards coherent control of an InAs quantum dot ensemble \cite{Suzukietal2016} and to resolve the hyperfine structure of rubidium in the presence of Doppler broadening \cite{LomsadzeCundiff2017}. However, MDCS studies on CQDs are scarce and have been primarily performed at room-temperature \cite{Turneretal2011,Carametal,Wongetal,Blocketal2012}.

In this Letter, we apply MDCS at cryogenic temperatures to study coherent dynamics of CdSe CQDs on the femtosecond timescale. Spectrally separating the third-order responses that involve intraband coherences from those that involve intermediate population states reveals differing behaviors as a function of excitation inter-pulse time delay. Comparison to simulation provides further evidence that strong modification of the exciton dephasing occurs via coupling to LO vibrational modes. Vibrational coupling, previously absent in discussions of spectral diffusion in CdSe CQDs, is found to play a primary role in the femtosecond timescale energy fluctuations of the exciton resonances that comprise one of the microscopic origins of dephasing in the material.

MDCS records a transient four-wave-mixing signal generated by three incident laser pulses as a function of two inter-pulse delays ($\tau$ and $T$) and evolution time ($t$) after the third pulse. The coherences excited by each pulse and their evolution frequencies are spectrally resolved and correlated by Fourier transforming the FWM signal along their respective time axes. Most commonly, the transformed time variables are $\tau$ and $t$, which results in a one-quantum spectrum with the conjugate axis to $\tau$ representing the absorption frequency $\omega_\tau$ and the conjugate axis to $t$ representing emission frequency $\omega_t$. The coherence generated by each first pulse evolves with negative phase for the rephasing signals \cite{HammZanni} measured here, which is reflected in negative values of $\omega_\tau$. In this study, we also record the signal as a function of delay $T$ and transform with respect to the variables $T$ and $t$, which generates a zero-quantum spectrum with the same emission frequency axis and the conjugate axis of $T$ representing the intraband coherence mixing frequency \cite{Yangetal2008} $\omega_T$.

We use a Multi-Dimensional Optical Nonlinear Spectrometer (MONSTR) \cite{Bristowetal2009}, which splits pulses from an optical parametric amplifier into four identical copies that are independently delayed in time and arranged in the box geometry \cite{Eckbreth1978}. The excitation pulses are of 90 fs duration at a 250 kHz repetition rate, and the excitation intensity of 4 W/cm$^2$ generates a predominately third-order response as verified by a power-dependence measurement of the heterodyned signal.  All pulses are co-linearly polarized and centered at wavelength $\lambda \approx 605$ nm. The sample is CdSe/CdZnS core/shell CQDs of 2 nm core radius and 2.5 nm shell thickness suspended in heptomethylnonane whose synthesis procedure is detailed elsewhere \cite{Limetal2014}. The sample optical density is 0.3 at the room-temperature 1S exciton absorption peak.

\begin{figure} 
  \includegraphics[width=0.45\textwidth]{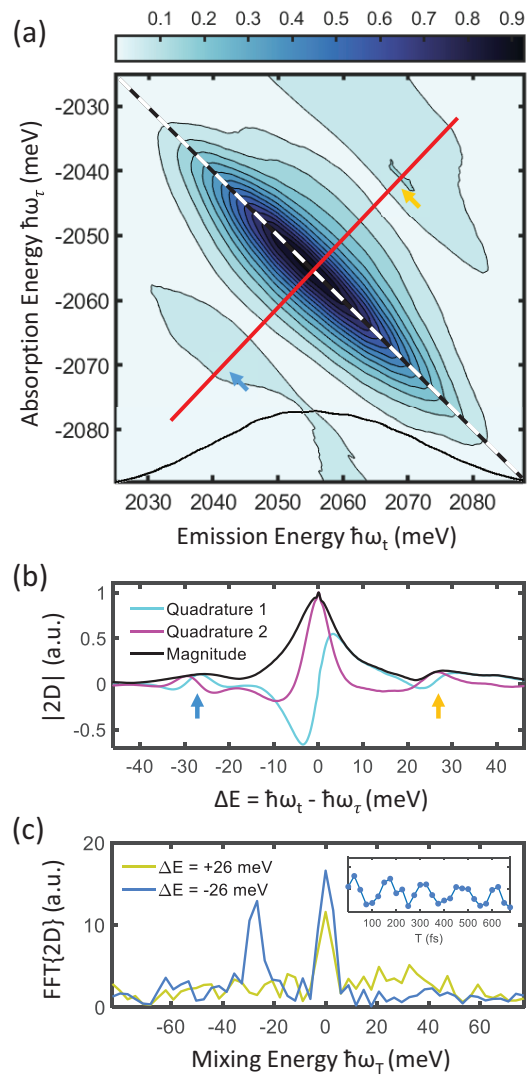}
  \caption{(a) Magnitude one-quantum spectrum at $T$ = 0. The dashed line and solid red line indicate the plot diagonal ($|\hbar\omega_\tau| = |\hbar\omega_t|$) and plot slice location respectively. The solid black curve represents the spectrum of the excitation and local oscillator pulses. (b) Magnitude and quadratures of the $T$ = 0 plot slice centered at $\hbar\omega_\tau = \hbar\omega_t = 2055$ meV. (c) Fourier transforms of the (twice zero-padded) complex evolutions of the $\Delta E = -26$ meV and its conjugate $\Delta E = +26$ meV point. These slice positions are marked by arrows in (a) and (b). Inset shows absolute value evolution of the $\Delta E = -26$ meV point.}
  \label{OneQuantumFigure}
\end{figure}

One-quantum spectra were acquired at a temperature of 16 K for delay $T$ increasing from 0 fs to 675 fs at 25 fs intervals. All spectra were phased relative to each other by maximizing the absorptive lineshape for one of the quadratures. As an example, we plot in Figs. \ref{OneQuantumFigure}(a) and \ref{OneQuantumFigure}(b) the one-quantum spectrum at $T$ = 0 and a slice perpendicular to the diagonal line. Two prominent features of the spectrum are a zero-phonon line at $\Delta E = E_t - E_\tau = 0$ and a surrounding broad pedestal at $|\Delta E| < 10$ meV due to coupling with the lattice acoustic phonon modes, which we will discuss in a future paper. Here, we focus on sidebands observed at energies $\Delta E \approx \pm$ 26 meV (matching the longitudinal-optical (LO) phonon mode energy $\hbar\omega_{LO}$ of CdSe \cite{Ferneeetal2014-2}), which are highlighted by the blue and yellow arrows in Fig. \ref{OneQuantumFigure}(a) and (b). In Fig. \ref{OneQuantumFigure}(c) it can be seen that Fourier transforming the evolution along time $T$ of the complex slices at $\Delta E = -26$ meV reveals a clear peak indicative of quantum oscillations in time $T$ corresponding to allowed intraband coherences at the LO-phonon energy. Such oscillations have previously been observed in integrated FWM signals by three-pulse FWM experiments \cite{Mittlemanetal1994}, but were not spectrally resolved and correlated in their absorption and emission dynamics. 

\begin{figure}[t] 
    \centering
    \includegraphics[width=0.5\textwidth]{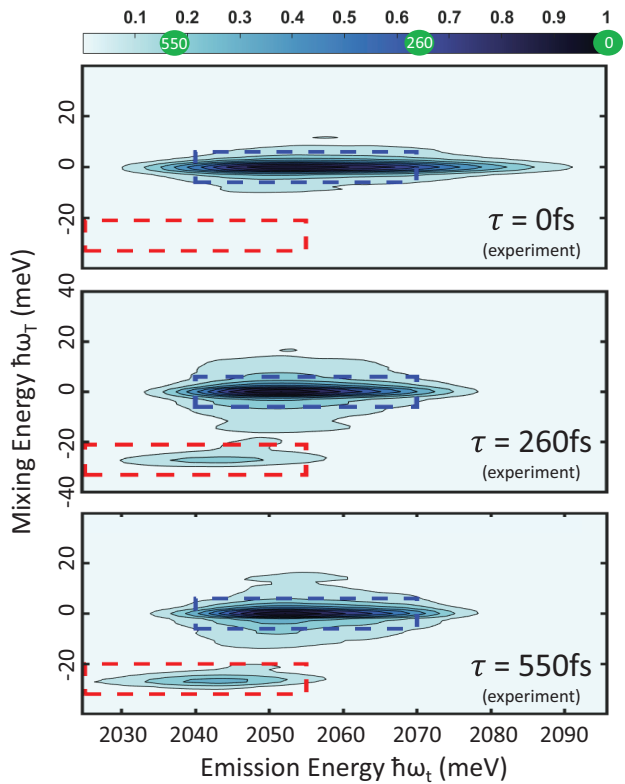}
    \caption{Zero-Quantum plots at $\tau$ = 0fs, 260fs, and 550fs as indicated. The dashed blue and red boxes indicate the areas integrated over to find their respective peak intensities. The relative normalization factors of each plot are indicated on the colorbar for reference.}
    \label{ZeroQuantumData}
\end{figure}

\begin{figure*} 
    \centering
    \includegraphics[width=1\textwidth]{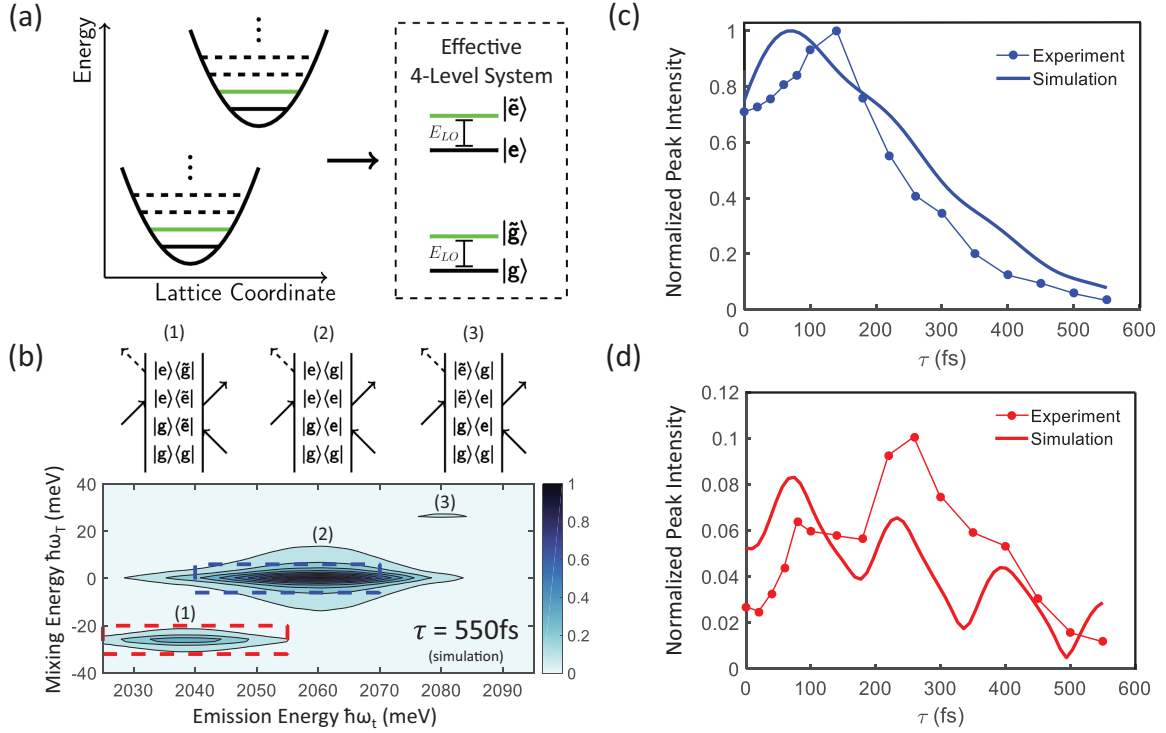}
    \caption{(a) Schematic of the reduced 4-level system which we use to interpret our data. (b) Simulated zero-quantum spsectrum at $\tau = 550$ fs with the parameters $S = 0.3$, $\tau_c = \tau_c^{vib} = 1$ ps, $\Delta\omega = 15$ meV, and $\Delta\omega^{vib} = 3$ meV. Three example double-sided Feynman diagrams are shown and their zero-quantum response positions $\{E_{emi},E_{mix}\}$ are indicated on the plot. The positions of the diagrams, labeled (1), (2), and (3), are $\{E_g - E_{LO},-E_{LO}\}$, $\{E_g,0\}$, and $\{E_g + E_{LO},+E_{LO}\}$ respectively. (c),(d) Evolution of the $E_{mix} = 0$ and $E_{mix} = -E_{LO}$ peak intensities (integrated over the colored boxed areas in (b)) respectively as a function of delay $\tau$. Both curves are normalized to the curve shown in (c).}
    \label{ZeroQuantumSimulation}
\end{figure*}

These one-quantum data reveal two main non-intuitive observations: (1) only the Stokes sideband in the one-quantum spectra exhibits strong oscillations due to the LO-phonon coupling as a function of $T$ and (2) its Fourier spectrum shown in Fig. \ref{OneQuantumFigure}(c) is one-sided. Understanding the origin of these observations will give further insight into the fundamental physical processes in CQDs.
Complicating the study of these one-quantum data however, is that the responses involving intraband coherences during $T$ appear at the same coordinates as those involving population states during $T$. A demonstrated way of separating overlapping pathways on a one-quantum spectrum is by selective excitation of certain response pathways via spectral filtering of the excitation pulses \cite{Tollerudetal2014,WenandNelson2013,Senliketal}. Another method is to acquire zero-quantum spectra, which spectrally separate the intraband coherence pathway responses from that of their population state counterparts directly \cite{Yangetal2008}. We thus acquire zero-quantum spectra at $\tau$ delays spanning 0 fs to 550 fs. Three representative zero-quantum spectra are plotted in Fig. \ref{ZeroQuantumData}. From these plots, it is clear that as $\tau$ increases, a sideband appears at $\omega_T = -26$ meV, the LO-phonon energy. There is also an asymmetry in the sideband formation as no apparent sideband is observed at $\omega_T = +26$ meV, in agreement with the one-sided spectrum plotted in Fig. \ref{OneQuantumFigure}(c). Most interestingly, integrating the spectrum over the blue and red dashed rectangles shown in Fig. \ref{ZeroQuantumData} and \ref{ZeroQuantumSimulation}(b) reveals that both peaks strengthen during early times $\tau$ (130 fs for the $\omega_T = 0$ peak and 250 fs for the sideband). The full evolutions are shown in Fig. \ref{ZeroQuantumSimulation}(c) and \ref{ZeroQuantumSimulation}(d). To explain these observations, we simulate the system's response and its resultant zero-quantum spectra. 

Motivated by the sideband positions observed in both one- and zero-quantum data, simulations are performed for a level system consisting of Franck-Condon transitions between ground and excited state manifolds formed from ladders of states separated by the LO-phonon energy \cite{MayandKuhn}. The oscillator strengths between states are proportional to their respective Franck-Condon factors \cite{deJongetal},
which are functions of the Huang-Rhys parameter $S$ (characterizing the electronic-vibrational coupling strength) and the initial/final vibrational excitation number $m$/$n$. Due to limited laser bandwidth of 30 meV and decreasing transition strength with higher $m$ and $n$, it is assumed that the main transitions contributing to the signal occur between the zeroth and first vibrational states in the ground $\{\Ket{g}, \Ket{\tilde{g}}\}$ and excited state manifolds $\{\Ket{e}, \Ket{\tilde{e}}\}$. We therefore reduce the dynamics of our system to that of a four-level system, as shown in Fig. \ref{ZeroQuantumSimulation}(a), and the ensemble-averaged transitions between these ladder states form the peaks in the simulated zero-quantum spectrum shown in Fig. \ref{ZeroQuantumSimulation}(b). We further simplify our simulation in two ways. First, since the sample temperature of 16 Kelvin is much lower than the LO-phonon Boltzmann temperature of 302 Kelvin, we assume all excited CQDs begin in the ground state $\Ket{g}$. Second, we repeated the zero-quantum experiment with co-circularly and cross-circularly polarized excitation beams and observed the same peak behaviors. Because the CQD selection rules dictate enhancement and suppression of doubly-excited transitions by cross-circular and co-circular excitation respectively \cite{Scholes2004}, we neglect transitions into doubly-excited states in the simulations.

To associate the observed peaks with evolution of excited coherences and populations, double-sided Feynman diagrams are used, which represent the quantum pathways that compose the system's perturbative response \cite{Mukamel1999}. The signal in the phase-matched direction that we measure is generated as follows: (1) The first pulse generates an interband coherence that evolves during delay $\tau$ at an energy within the laser spectrum, called a one-quantum coherence.  (2) The second pulse generates either a population state or an intraband coherence that evolves during delay $T$ at an energy within the laser bandwidth, called a zero-quantum coherence. (3) The third pulse generates the last interband (one-quantum) coherence that radiates as a coherent FWM signal during $t$. The changes induced in the system's density matrix by each pulse are collectively known as Liouville pathways, and a comprehensive explanation of how the pathways represented by each diagram map onto the various peaks of a zero-quantum spectrum is given by Yang et al. \cite{Yangetal2008}. For brevity we show in Fig. \ref{ZeroQuantumSimulation}(b) three example diagrams and the positions at which their responses will appear on the simulated zero-quantum spectra.

Liouville pathways associate each peak's rise in $\tau$ with the evolution of interband coherences generated by the first excitation pulse. Such rises may be due to time-ordering ambiguities at inter-pulse delays shorter than the pulse durations, but the 90 fs pulses used in this study are too short for coherent signals during pulse overlap to be the main cause. However, inclusion of non-Markovian dynamics allows for a photon echo integration \cite{LorenzandCundiff2005} rise to occur. Non-Markovian dephasing lineshapes are commonly obtained by applying the cumulant expansion to the spectral diffusion trajectory $\delta\omega_{ij}(t)$ of a coherence $\rho_{ij}$, where $i,j = \{g,e,\tilde{g},\tilde{e}\}$, and truncating at second-order \cite{Mukamel1999}:
\begin{align}
    \rho_{ij}(t) &\propto e^{-i\omega_{ij}t}\Braket{e^{-i\int^t_0\delta\omega_{ij}(\tau)d\tau}} \approx e^{-i\omega_{ij}t}e^{-g(t)}
\end{align}
where the lineshape function $g(t)$ is determined by the two-point correlation function $C(t) = \Braket{\delta\omega_{ij}(t)\omega_{ij}(0)}$ of the transition \cite{Mukamel1999}. Here, we use the simple lineshape function resulting from the exponentially decaying Kubo ansatz \cite{HammZanni} $C(t) = \Delta\omega^2e^{-\frac{|t|}{\tau_c}}$, where $\Delta\omega$ and $\tau_c$ are the amplitude and correlation time of the spectral diffusion respectively.

A zero-quantum spectrum simulated for the Kubo dephasing lineshape is plotted in Fig. \ref{ZeroQuantumSimulation}(b). Comparison between the experimental and simulated spectra at $\tau = 550$ fs shows very good agreement between the peak positions and relative intensities. Crucially, we achieve this agreement by assigning a large spectral diffusion amplitude $\Delta\omega = 15$ meV to the ``vibration-less coherences" ($\rho_{eg}$ and $\rho_{ge}$) and a comparatively smaller amplitude $\Delta\omega^{vib}$ = 3 meV to ``coupled coherences" (all $\rho_{ij}$ involving states $\tilde{g}$ and $\tilde{e}$). Because no sidebands appear in simulation if $\Delta\omega = \Delta\omega^{vib}$, the sideband observed in experiment indicates strong modification of coherence dephasing dynamics via coupling to lattice LO vibrational modes. However, Fig. \ref{ZeroQuantumSimulation}(d) shows that choosing dephasing parameters for matching the decay rate at large $\tau$ results in sideband rise times much shorter than the 250 fs rise time observed from experiment. This discrepancy between experiment and simulation for the sideband evolution indicates that coupled coherences in our system exhibit non-Markovian dynamics beyond the Kubo ansatz or even the usual second-order cumulant truncation \cite{Chenetal2015}.

Oscillations at  $\omega_{LO}$ are also observed in the sideband evolution for both the experimental and simulated curves shown in Fig. \ref{ZeroQuantumSimulation}(d). Beating due to a coherence evolution cannot be the origin because any quantum coherences excited during $\tau$ must have an energy difference at the excitation frequency. These oscillations are thus due to polarization interference \cite{MeierThomasKoch} between Liouville pathways with initial interband coherences differing in energy by $\hbar\omega_{LO}$. Specifically in our model, the interference is between the ground state bleach pathways that access each of the two ground state vibrational levels. The phases of these oscillations in the simulations and data are matched by adding a phase of $\frac{4}{5}\pi$ to the simulated coupled coherence responses.

To date, two regimes of spectral diffusion have been identified, on the seconds timescale \cite{Ferneeetal2012} and sub-20 $\mu$s timescale \cite{Coolenetal2008}. Previous studies have focused on free surface charges \cite{Mulleretal2004} and surface ligand rearrangement \cite{Ferneeetal2012} as possible causes for the band-edge Stark shift \cite{EmpedoclesandBawendi1997} that leads to spectral diffusion, and a model has been proposed by Beyler et al. \cite{Beyleretal2013} that attributes spectral diffusion at all timescales to cooperative effects of the above phenomena. The above theories are not sufficient to explain our results, which clearly point to coupling with lattice LO vibrational modes as a major factor in the spectral diffusion dynamics on femtosecond timescales. We suspect that the random environmental perturbations that contribute to the energy gap fluctuations become less dominant when nuclear motion is initiated in the LO vibrational mode. The local fields induced by the nuclear motion extend over many unit cells - effectively over the entire core volume of our CQDs. For CQDs grown with a thick shell structure, such as for our sample, it is reasonable to expect that surface charge dynamics are a weak effect compared to the Frohlich coupling between the exciton and local fields that synchronizes the exciton motion with that of the entire CQD core lattice. The spectral diffusion dynamics may then approach timescales on the order of the LO phonon period $T_{LO} \approx 150$ fs, and highly non-Markovian evolution of the coupled coherences could then occur. A more detailed microscopic theory is needed to elucidate the exact mechanisms responsible.

In conclusion, we have found that spectral diffusion of the exciton resonances in CdSe CQDs is strongly modified in the presence of coupled vibrational excitations. These results will prove crucial in applications of CQDs and other systems with strong electronic-vibrational coupling towards areas in which pure decoherence is a primary obstacle (e.g. single-photon emitter \cite{Linetal2017,Chandrasekaranetal2017} and quantum information \cite{PryorandFlatte2006,Engeletal2001}). Furthermore, non-Markovian effects, previously thought to be purely detrimental, have been shown in recent years to have possible benefits towards implementing quantum information protocols \cite{Liuetal2013,Laineetal2014}. Further theoretical investigation into vibrational modification of spectral diffusion in CQDs and its microscopic origin is thus warranted for understanding its implications in CQD device design and their fundamental properties.

\begin{acknowledgments}
	This work was supported by the Department of Energy grant number DE-SC0015782. D.B.A. acknowledges support by a fellowship from the Brazilian National Council for Scientific and Technological Development (CNPq). L.A.P. acknowledges support from FAPESP (Project numbers 2013/16911-2 and 2016/50011-7).
\end{acknowledgments}


\begin{thebibliography}{40}
\expandafter\ifx\csname natexlab\endcsname\relax\def\natexlab#1{#1}\fi
\expandafter\ifx\csname bibnamefont\endcsname\relax
  \def\bibnamefont#1{#1}\fi
\expandafter\ifx\csname bibfnamefont\endcsname\relax
  \def\bibfnamefont#1{#1}\fi
\expandafter\ifx\csname citenamefont\endcsname\relax
  \def\citenamefont#1{#1}\fi
\expandafter\ifx\csname url\endcsname\relax
  \def\url#1{\texttt{#1}}\fi
\expandafter\ifx\csname urlprefix\endcsname\relax\def\urlprefix{URL }\fi
\providecommand{\bibinfo}[2]{#2}
\providecommand{\eprint}[2][]{\url{#2}}

\bibitem[{\citenamefont{Medintz et~al.}(2005)\citenamefont{Medintz, Uyeda,
  Goldman, and Mattoussi}}]{Medintzetal2005}
\bibinfo{author}{\bibfnamefont{I.~L.} \bibnamefont{Medintz}},
  \bibinfo{author}{\bibfnamefont{H.~T.} \bibnamefont{Uyeda}},
  \bibinfo{author}{\bibfnamefont{E.~R.} \bibnamefont{Goldman}},
  \bibnamefont{and}
  \bibinfo{author}{\bibfnamefont{H.}~\bibnamefont{Mattoussi}},
  \bibinfo{journal}{Nature Materials} \textbf{\bibinfo{volume}{4}},
  \bibinfo{pages}{435 EP } (\bibinfo{year}{2005}), \bibinfo{note}{review
  Article}, \urlprefix\url{http://dx.doi.org/10.1038/nmat1390}.

\bibitem[{\citenamefont{Eunjoo et~al.}(2010)\citenamefont{Eunjoo, Shinae,
  Hyosook, Jungeun, Byungki, and Younghwan}}]{Jangetal2010}
\bibinfo{author}{\bibfnamefont{J.}~\bibnamefont{Eunjoo}},
  \bibinfo{author}{\bibfnamefont{J.}~\bibnamefont{Shinae}},
  \bibinfo{author}{\bibfnamefont{J.}~\bibnamefont{Hyosook}},
  \bibinfo{author}{\bibfnamefont{L.}~\bibnamefont{Jungeun}},
  \bibinfo{author}{\bibfnamefont{K.}~\bibnamefont{Byungki}}, \bibnamefont{and}
  \bibinfo{author}{\bibfnamefont{K.}~\bibnamefont{Younghwan}},
  \bibinfo{journal}{Advanced Materials} \textbf{\bibinfo{volume}{22}},
  \bibinfo{pages}{3076} (\bibinfo{year}{2010}), ISSN \bibinfo{issn}{0935-9648},
  \urlprefix\url{https://doi.org/10.1002/adma.201000525}.

\bibitem[{\citenamefont{Lan et~al.}(2014)\citenamefont{Lan, Masala, and
  Sargent}}]{Lanetal2014}
\bibinfo{author}{\bibfnamefont{X.}~\bibnamefont{Lan}},
  \bibinfo{author}{\bibfnamefont{S.}~\bibnamefont{Masala}}, \bibnamefont{and}
  \bibinfo{author}{\bibfnamefont{E.~H.} \bibnamefont{Sargent}},
  \bibinfo{journal}{Nature Materials} \textbf{\bibinfo{volume}{13}},
  \bibinfo{pages}{233 EP } (\bibinfo{year}{2014}),
  \urlprefix\url{http://dx.doi.org/10.1038/nmat3816}.

\bibitem[{\citenamefont{Fernee et~al.}(2014{\natexlab{a}})\citenamefont{Fernee,
  Tamarat, and Lounis}}]{Ferneeetal2014-1}
\bibinfo{author}{\bibfnamefont{M.~J.} \bibnamefont{Fernee}},
  \bibinfo{author}{\bibfnamefont{P.}~\bibnamefont{Tamarat}}, \bibnamefont{and}
  \bibinfo{author}{\bibfnamefont{B.}~\bibnamefont{Lounis}},
  \bibinfo{journal}{Chemical Society Reviews} \textbf{\bibinfo{volume}{43}},
  \bibinfo{pages}{1311} (\bibinfo{year}{2014}{\natexlab{a}}), ISSN
  \bibinfo{issn}{0306-0012},
  \urlprefix\url{http://dx.doi.org/10.1039/C3CS60209E}.

\bibitem[{\citenamefont{Coolen et~al.}(2008)\citenamefont{Coolen, Brokmann,
  Spinicelli, and Hermier}}]{Coolenetal2008}
\bibinfo{author}{\bibfnamefont{L.}~\bibnamefont{Coolen}},
  \bibinfo{author}{\bibfnamefont{X.}~\bibnamefont{Brokmann}},
  \bibinfo{author}{\bibfnamefont{P.}~\bibnamefont{Spinicelli}},
  \bibnamefont{and} \bibinfo{author}{\bibfnamefont{J.-P.}
  \bibnamefont{Hermier}}, \bibinfo{journal}{Physical Review Letters}
  \textbf{\bibinfo{volume}{100}}, \bibinfo{pages}{027403}
  (\bibinfo{year}{2008}),
  \urlprefix\url{https://link.aps.org/doi/10.1103/PhysRevLett.100.027403}.

\bibitem[{\citenamefont{Beyler et~al.}(2013)\citenamefont{Beyler, Marshall,
  Cui, Brokmann, and Bawendi}}]{Beyleretal2013}
\bibinfo{author}{\bibfnamefont{A.~P.} \bibnamefont{Beyler}},
  \bibinfo{author}{\bibfnamefont{L.~F.} \bibnamefont{Marshall}},
  \bibinfo{author}{\bibfnamefont{J.}~\bibnamefont{Cui}},
  \bibinfo{author}{\bibfnamefont{X.}~\bibnamefont{Brokmann}}, \bibnamefont{and}
  \bibinfo{author}{\bibfnamefont{M.~G.} \bibnamefont{Bawendi}},
  \bibinfo{journal}{Physical Review Letters} \textbf{\bibinfo{volume}{111}},
  \bibinfo{pages}{177401} (\bibinfo{year}{2013}),
  \urlprefix\url{https://link.aps.org/doi/10.1103/PhysRevLett.111.177401}.

\bibitem[{\citenamefont{Cundiff and Mukamel}(2013)}]{CundiffMukamel2013}
\bibinfo{author}{\bibfnamefont{S.~T.} \bibnamefont{Cundiff}} \bibnamefont{and}
  \bibinfo{author}{\bibfnamefont{S.}~\bibnamefont{Mukamel}},
  \bibinfo{journal}{Physics Today} \textbf{\bibinfo{volume}{66}},
  \bibinfo{pages}{44} (\bibinfo{year}{2013}), ISSN \bibinfo{issn}{0031-9228},
  \urlprefix\url{https://doi.org/10.1063/PT.3.2047}.

\bibitem[{\citenamefont{Ferrio and Steel}(1998)}]{FerrioSteel1998}
\bibinfo{author}{\bibfnamefont{K.~B.} \bibnamefont{Ferrio}} \bibnamefont{and}
  \bibinfo{author}{\bibfnamefont{D.~G.} \bibnamefont{Steel}},
  \bibinfo{journal}{Physical Review Letters} \textbf{\bibinfo{volume}{80}},
  \bibinfo{pages}{786} (\bibinfo{year}{1998}),
  \urlprefix\url{https://link.aps.org/doi/10.1103/PhysRevLett.80.786}.

\bibitem[{\citenamefont{Suzuki et~al.}(2016)\citenamefont{Suzuki, Singh, Bayer,
  Ludwig, Wieck, and Cundiff}}]{Suzukietal2016}
\bibinfo{author}{\bibfnamefont{T.}~\bibnamefont{Suzuki}},
  \bibinfo{author}{\bibfnamefont{R.}~\bibnamefont{Singh}},
  \bibinfo{author}{\bibfnamefont{M.}~\bibnamefont{Bayer}},
  \bibinfo{author}{\bibfnamefont{A.}~\bibnamefont{Ludwig}},
  \bibinfo{author}{\bibfnamefont{A.~D.} \bibnamefont{Wieck}}, \bibnamefont{and}
  \bibinfo{author}{\bibfnamefont{S.~T.} \bibnamefont{Cundiff}},
  \bibinfo{journal}{Physical Review Letters} \textbf{\bibinfo{volume}{117}},
  \bibinfo{pages}{157402} (\bibinfo{year}{2016}),
  \urlprefix\url{https://link.aps.org/doi/10.1103/PhysRevLett.117.157402}.

\bibitem[{\citenamefont{Lomsadze and Cundiff}(2017)}]{LomsadzeCundiff2017}
\bibinfo{author}{\bibfnamefont{B.}~\bibnamefont{Lomsadze}} \bibnamefont{and}
  \bibinfo{author}{\bibfnamefont{S.~T.} \bibnamefont{Cundiff}},
  \bibinfo{journal}{Science} \textbf{\bibinfo{volume}{357}},
  \bibinfo{pages}{1389} (\bibinfo{year}{2017}),
  \urlprefix\url{http://science.sciencemag.org/content/357/6358/1389.abstract}.

\bibitem[{\citenamefont{Turner et~al.}(2012)\citenamefont{Turner, Hassan, and
  Scholes}}]{Turneretal2011}
\bibinfo{author}{\bibfnamefont{D.~B.} \bibnamefont{Turner}},
  \bibinfo{author}{\bibfnamefont{Y.}~\bibnamefont{Hassan}}, \bibnamefont{and}
  \bibinfo{author}{\bibfnamefont{G.~D.} \bibnamefont{Scholes}},
  \bibinfo{journal}{Nano Letters} \textbf{\bibinfo{volume}{12}},
  \bibinfo{pages}{880} (\bibinfo{year}{2012}), ISSN \bibinfo{issn}{1530-6984},
  \urlprefix\url{https://doi.org/10.1021/nl2039502}.

\bibitem[{\citenamefont{Caram et~al.}(2014)\citenamefont{Caram, Zheng,
  Dahlberg, Rolczynski, Griffin, Fidler, Dolzhnikov, Talapin, and
  Engel}}]{Carametal}
\bibinfo{author}{\bibfnamefont{J.~R.} \bibnamefont{Caram}},
  \bibinfo{author}{\bibfnamefont{H.}~\bibnamefont{Zheng}},
  \bibinfo{author}{\bibfnamefont{P.~D.} \bibnamefont{Dahlberg}},
  \bibinfo{author}{\bibfnamefont{B.~S.} \bibnamefont{Rolczynski}},
  \bibinfo{author}{\bibfnamefont{G.~B.} \bibnamefont{Griffin}},
  \bibinfo{author}{\bibfnamefont{A.~F.} \bibnamefont{Fidler}},
  \bibinfo{author}{\bibfnamefont{D.~S.} \bibnamefont{Dolzhnikov}},
  \bibinfo{author}{\bibfnamefont{D.~V.} \bibnamefont{Talapin}},
  \bibnamefont{and} \bibinfo{author}{\bibfnamefont{G.~S.} \bibnamefont{Engel}},
  \bibinfo{journal}{The Journal of Physical Chemistry Letters}
  \textbf{\bibinfo{volume}{5}}, \bibinfo{pages}{196} (\bibinfo{year}{2014}),
  ISSN \bibinfo{issn}{1948-7185},
  \urlprefix\url{https://doi.org/10.1021/jz402336t}.

\bibitem[{\citenamefont{Wong and Scholes}(2011)}]{Wongetal}
\bibinfo{author}{\bibfnamefont{C.~Y.} \bibnamefont{Wong}} \bibnamefont{and}
  \bibinfo{author}{\bibfnamefont{G.~D.} \bibnamefont{Scholes}},
  \bibinfo{journal}{The Journal of Physical Chemistry A}
  \textbf{\bibinfo{volume}{115}}, \bibinfo{pages}{3797} (\bibinfo{year}{2011}),
  ISSN \bibinfo{issn}{1089-5639},
  \urlprefix\url{https://doi.org/10.1021/jp1079197}.

\bibitem[{\citenamefont{Block et~al.}(2012)\citenamefont{Block, Yurs, Pakoulev,
  Selinsky, Jin, and Wright}}]{Blocketal2012}
\bibinfo{author}{\bibfnamefont{S.~B.} \bibnamefont{Block}},
  \bibinfo{author}{\bibfnamefont{L.~A.} \bibnamefont{Yurs}},
  \bibinfo{author}{\bibfnamefont{A.~V.} \bibnamefont{Pakoulev}},
  \bibinfo{author}{\bibfnamefont{R.~S.} \bibnamefont{Selinsky}},
  \bibinfo{author}{\bibfnamefont{S.}~\bibnamefont{Jin}}, \bibnamefont{and}
  \bibinfo{author}{\bibfnamefont{J.~C.} \bibnamefont{Wright}},
  \bibinfo{journal}{The Journal of Physical Chemistry Letters}
  \textbf{\bibinfo{volume}{3}}, \bibinfo{pages}{2707} (\bibinfo{year}{2012}),
  ISSN \bibinfo{issn}{1948-7185},
  \urlprefix\url{https://doi.org/10.1021/jz300599b}.

\bibitem[{\citenamefont{Hamm and Zanni}(2011)}]{HammZanni}
\bibinfo{author}{\bibfnamefont{P.}~\bibnamefont{Hamm}} \bibnamefont{and}
  \bibinfo{author}{\bibfnamefont{M.}~\bibnamefont{Zanni}},
  \emph{\bibinfo{title}{Concepts and Methods of 2D Infrared Spectroscopy}}
  (\bibinfo{publisher}{Cambridge University Press}, \bibinfo{year}{2011}),
  \bibinfo{edition}{1st} ed.

\bibitem[{\citenamefont{Yang et~al.}(2008)\citenamefont{Yang, Zhang, Bristow,
  Cundiff, and Mukamel}}]{Yangetal2008}
\bibinfo{author}{\bibfnamefont{L.}~\bibnamefont{Yang}},
  \bibinfo{author}{\bibfnamefont{T.}~\bibnamefont{Zhang}},
  \bibinfo{author}{\bibfnamefont{A.~D.} \bibnamefont{Bristow}},
  \bibinfo{author}{\bibfnamefont{S.~T.} \bibnamefont{Cundiff}},
  \bibnamefont{and} \bibinfo{author}{\bibfnamefont{S.}~\bibnamefont{Mukamel}},
  \bibinfo{journal}{The Journal of Chemical Physics}
  \textbf{\bibinfo{volume}{129}}, \bibinfo{pages}{234711}
  (\bibinfo{year}{2008}), ISSN \bibinfo{issn}{0021-9606},
  \urlprefix\url{https://doi.org/10.1063/1.3037217}.

\bibitem[{\citenamefont{Bristow et~al.}(2009)\citenamefont{Bristow, Karaiskaj,
  Dai, Zhang, Carlsson, Hagen, Jimenez, and Cundiff}}]{Bristowetal2009}
\bibinfo{author}{\bibfnamefont{A.~D.} \bibnamefont{Bristow}},
  \bibinfo{author}{\bibfnamefont{D.}~\bibnamefont{Karaiskaj}},
  \bibinfo{author}{\bibfnamefont{X.}~\bibnamefont{Dai}},
  \bibinfo{author}{\bibfnamefont{T.}~\bibnamefont{Zhang}},
  \bibinfo{author}{\bibfnamefont{C.}~\bibnamefont{Carlsson}},
  \bibinfo{author}{\bibfnamefont{K.~R.} \bibnamefont{Hagen}},
  \bibinfo{author}{\bibfnamefont{R.}~\bibnamefont{Jimenez}}, \bibnamefont{and}
  \bibinfo{author}{\bibfnamefont{S.~T.} \bibnamefont{Cundiff}},
  \bibinfo{journal}{Review of Scientific Instruments}
  \textbf{\bibinfo{volume}{80}}, \bibinfo{pages}{073108}
  (\bibinfo{year}{2009}), ISSN \bibinfo{issn}{0034-6748},
  \urlprefix\url{https://doi.org/10.1063/1.3184103}.

\bibitem[{\citenamefont{Eckbreth}(1978)}]{Eckbreth1978}
\bibinfo{author}{\bibfnamefont{A.~C.} \bibnamefont{Eckbreth}},
  \bibinfo{journal}{Applied Physics Letters} \textbf{\bibinfo{volume}{32}},
  \bibinfo{pages}{421} (\bibinfo{year}{1978}), ISSN \bibinfo{issn}{0003-6951},
  \urlprefix\url{https://doi.org/10.1063/1.90070}.

\bibitem[{\citenamefont{Jaehoon et~al.}(2014)\citenamefont{Jaehoon, Guk,
  Myeongjin, Kyeong, M., Young-Shin, I., Changhee, C., and Ki}}]{Limetal2014}
\bibinfo{author}{\bibfnamefont{L.}~\bibnamefont{Jaehoon}},
  \bibinfo{author}{\bibfnamefont{J.~B.} \bibnamefont{Guk}},
  \bibinfo{author}{\bibfnamefont{P.}~\bibnamefont{Myeongjin}},
  \bibinfo{author}{\bibfnamefont{K.~J.} \bibnamefont{Kyeong}},
  \bibinfo{author}{\bibfnamefont{P.~J.} \bibnamefont{M.}},
  \bibinfo{author}{\bibfnamefont{P.}~\bibnamefont{Young-Shin}},
  \bibinfo{author}{\bibfnamefont{K.~V.} \bibnamefont{I.}},
  \bibinfo{author}{\bibfnamefont{L.}~\bibnamefont{Changhee}},
  \bibinfo{author}{\bibfnamefont{L.~D.} \bibnamefont{C.}}, \bibnamefont{and}
  \bibinfo{author}{\bibfnamefont{B.~W.} \bibnamefont{Ki}},
  \bibinfo{journal}{Advanced Materials} \textbf{\bibinfo{volume}{26}},
  \bibinfo{pages}{8034} (\bibinfo{year}{2014}), ISSN \bibinfo{issn}{0935-9648},
  \urlprefix\url{https://doi.org/10.1002/adma.201403620}.

\bibitem[{\citenamefont{Fernee et~al.}(2014{\natexlab{b}})\citenamefont{Fernee,
  Sinito, Mulvaney, Tamarat, and Lounis}}]{Ferneeetal2014-2}
\bibinfo{author}{\bibfnamefont{M.~J.} \bibnamefont{Fernee}},
  \bibinfo{author}{\bibfnamefont{C.}~\bibnamefont{Sinito}},
  \bibinfo{author}{\bibfnamefont{P.}~\bibnamefont{Mulvaney}},
  \bibinfo{author}{\bibfnamefont{P.}~\bibnamefont{Tamarat}}, \bibnamefont{and}
  \bibinfo{author}{\bibfnamefont{B.}~\bibnamefont{Lounis}},
  \bibinfo{journal}{Physical Chemistry Chemical Physics}
  \textbf{\bibinfo{volume}{16}}, \bibinfo{pages}{16957}
  (\bibinfo{year}{2014}{\natexlab{b}}), ISSN \bibinfo{issn}{1463-9076},
  \urlprefix\url{http://dx.doi.org/10.1039/C4CP02022G}.

\bibitem[{\citenamefont{Mittleman et~al.}(1994)\citenamefont{Mittleman,
  Schoenlein, Shiang, Colvin, Alivisatos, and Shank}}]{Mittlemanetal1994}
\bibinfo{author}{\bibfnamefont{D.~M.} \bibnamefont{Mittleman}},
  \bibinfo{author}{\bibfnamefont{R.~W.} \bibnamefont{Schoenlein}},
  \bibinfo{author}{\bibfnamefont{J.~J.} \bibnamefont{Shiang}},
  \bibinfo{author}{\bibfnamefont{V.~L.} \bibnamefont{Colvin}},
  \bibinfo{author}{\bibfnamefont{A.~P.} \bibnamefont{Alivisatos}},
  \bibnamefont{and} \bibinfo{author}{\bibfnamefont{C.~V.} \bibnamefont{Shank}},
  \bibinfo{journal}{Physical Review B} \textbf{\bibinfo{volume}{49}},
  \bibinfo{pages}{14435} (\bibinfo{year}{1994}),
  \urlprefix\url{https://link.aps.org/doi/10.1103/PhysRevB.49.14435}.

\bibitem[{\citenamefont{Tollerud et~al.}(2014)\citenamefont{Tollerud, Hall, and
  Davis}}]{Tollerudetal2014}
\bibinfo{author}{\bibfnamefont{J.~O.} \bibnamefont{Tollerud}},
  \bibinfo{author}{\bibfnamefont{C.~R.} \bibnamefont{Hall}}, \bibnamefont{and}
  \bibinfo{author}{\bibfnamefont{J.~A.} \bibnamefont{Davis}},
  \bibinfo{journal}{Optics Express} \textbf{\bibinfo{volume}{22}},
  \bibinfo{pages}{6719} (\bibinfo{year}{2014}),
  \urlprefix\url{http://www.opticsexpress.org/abstract.cfm?URI=oe-22-6-6719}.

\bibitem[{\citenamefont{Wen and Nelson}(2013)}]{WenandNelson2013}
\bibinfo{author}{\bibfnamefont{P.}~\bibnamefont{Wen}} \bibnamefont{and}
  \bibinfo{author}{\bibfnamefont{K.~A.} \bibnamefont{Nelson}},
  \bibinfo{journal}{The Journal of Physical Chemistry A}
  \textbf{\bibinfo{volume}{117}}, \bibinfo{pages}{6380} (\bibinfo{year}{2013}),
  ISSN \bibinfo{issn}{1089-5639},
  \urlprefix\url{https://doi.org/10.1021/jp401150d}.

\bibitem[{\citenamefont{Senlik et~al.}(2015)\citenamefont{Senlik, Policht, and
  Ogilvie}}]{Senliketal}
\bibinfo{author}{\bibfnamefont{S.~S.} \bibnamefont{Senlik}},
  \bibinfo{author}{\bibfnamefont{V.~R.} \bibnamefont{Policht}},
  \bibnamefont{and} \bibinfo{author}{\bibfnamefont{J.~P.}
  \bibnamefont{Ogilvie}}, \bibinfo{journal}{The Journal of Physical Chemistry
  Letters} \textbf{\bibinfo{volume}{6}}, \bibinfo{pages}{2413}
  (\bibinfo{year}{2015}), ISSN \bibinfo{issn}{1948-7185},
  \urlprefix\url{https://doi.org/10.1021/acs.jpclett.5b00861}.

\bibitem[{\citenamefont{May and Kuhn}(2011)}]{MayandKuhn}
\bibinfo{author}{\bibfnamefont{V.}~\bibnamefont{May}} \bibnamefont{and}
  \bibinfo{author}{\bibfnamefont{O.}~\bibnamefont{Kuhn}},
  \emph{\bibinfo{title}{Charge and Energy Transfer Dynamics in Molecular
  Systems}} (\bibinfo{publisher}{Wiley-VCH Verlag}, \bibinfo{year}{2011}),
  \bibinfo{edition}{3rd} ed.

\bibitem[{\citenamefont{de~Jong et~al.}(2015)\citenamefont{de~Jong, Seijo,
  Meijerink, and Rabouw}}]{deJongetal}
\bibinfo{author}{\bibfnamefont{M.}~\bibnamefont{de~Jong}},
  \bibinfo{author}{\bibfnamefont{L.}~\bibnamefont{Seijo}},
  \bibinfo{author}{\bibfnamefont{A.}~\bibnamefont{Meijerink}},
  \bibnamefont{and} \bibinfo{author}{\bibfnamefont{F.~T.}
  \bibnamefont{Rabouw}}, \bibinfo{journal}{Physical Chemistry Chemical Physics}
  \textbf{\bibinfo{volume}{17}}, \bibinfo{pages}{16959} (\bibinfo{year}{2015}),
  ISSN \bibinfo{issn}{1463-9076},
  \urlprefix\url{http://dx.doi.org/10.1039/C5CP02093J}.

\bibitem[{\citenamefont{Scholes}(2004)}]{Scholes2004}
\bibinfo{author}{\bibfnamefont{G.~D.} \bibnamefont{Scholes}},
  \bibinfo{journal}{The Journal of Chemical Physics}
  \textbf{\bibinfo{volume}{121}}, \bibinfo{pages}{10104}
  (\bibinfo{year}{2004}), ISSN \bibinfo{issn}{0021-9606},
  \urlprefix\url{https://doi.org/10.1063/1.1808414}.

\bibitem[{\citenamefont{Mukamel}(1999)}]{Mukamel1999}
\bibinfo{author}{\bibfnamefont{S.}~\bibnamefont{Mukamel}},
  \emph{\bibinfo{title}{Principles of Nonlinear Optical Spectroscopy}}
  (\bibinfo{publisher}{Oxford University Press}, \bibinfo{year}{1999}),
  \bibinfo{edition}{1st} ed.

\bibitem[{\citenamefont{Lorenz and Cundiff}(2005)}]{LorenzandCundiff2005}
\bibinfo{author}{\bibfnamefont{V.~O.} \bibnamefont{Lorenz}} \bibnamefont{and}
  \bibinfo{author}{\bibfnamefont{S.~T.} \bibnamefont{Cundiff}},
  \bibinfo{journal}{Physical Review Letters} \textbf{\bibinfo{volume}{95}},
  \bibinfo{pages}{163601} (\bibinfo{year}{2005}),
  \urlprefix\url{https://link.aps.org/doi/10.1103/PhysRevLett.95.163601}.

\bibitem[{\citenamefont{Chen et~al.}(2015)\citenamefont{Chen, Lambert, Cheng,
  Chen, and Nori}}]{Chenetal2015}
\bibinfo{author}{\bibfnamefont{H.-B.} \bibnamefont{Chen}},
  \bibinfo{author}{\bibfnamefont{N.}~\bibnamefont{Lambert}},
  \bibinfo{author}{\bibfnamefont{Y.-C.} \bibnamefont{Cheng}},
  \bibinfo{author}{\bibfnamefont{Y.-N.} \bibnamefont{Chen}}, \bibnamefont{and}
  \bibinfo{author}{\bibfnamefont{F.}~\bibnamefont{Nori}},
  \bibinfo{journal}{Scientific Reports} \textbf{\bibinfo{volume}{5}},
  \bibinfo{pages}{12753 EP } (\bibinfo{year}{2015}), \bibinfo{note}{article},
  \urlprefix\url{http://dx.doi.org/10.1038/srep12753}.

\bibitem[{\citenamefont{Meier et~al.}(2006)\citenamefont{Meier, Thomas, and
  Koch}}]{MeierThomasKoch}
\bibinfo{author}{\bibfnamefont{T.}~\bibnamefont{Meier}},
  \bibinfo{author}{\bibfnamefont{P.}~\bibnamefont{Thomas}}, \bibnamefont{and}
  \bibinfo{author}{\bibfnamefont{S.}~\bibnamefont{Koch}},
  \emph{\bibinfo{title}{Coherent Semiconductor Optics: From Basic Concepts to
  Nanostructure Applications}} (\bibinfo{publisher}{Springer},
  \bibinfo{year}{2006}), \bibinfo{edition}{1st} ed.

\bibitem[{\citenamefont{Fern{\'e}e et~al.}(2012)\citenamefont{Fern{\'e}e,
  Plakhotnik, Louyer, Littleton, Potzner, Tamarat, Mulvaney, and
  Lounis}}]{Ferneeetal2012}
\bibinfo{author}{\bibfnamefont{M.~J.} \bibnamefont{Fern{\'e}e}},
  \bibinfo{author}{\bibfnamefont{T.}~\bibnamefont{Plakhotnik}},
  \bibinfo{author}{\bibfnamefont{Y.}~\bibnamefont{Louyer}},
  \bibinfo{author}{\bibfnamefont{B.~N.} \bibnamefont{Littleton}},
  \bibinfo{author}{\bibfnamefont{C.}~\bibnamefont{Potzner}},
  \bibinfo{author}{\bibfnamefont{P.}~\bibnamefont{Tamarat}},
  \bibinfo{author}{\bibfnamefont{P.}~\bibnamefont{Mulvaney}}, \bibnamefont{and}
  \bibinfo{author}{\bibfnamefont{B.}~\bibnamefont{Lounis}},
  \bibinfo{journal}{The Journal of Physical Chemistry Letters}
  \textbf{\bibinfo{volume}{3}}, \bibinfo{pages}{1716} (\bibinfo{year}{2012}),
  ISSN \bibinfo{issn}{1948-7185},
  \urlprefix\url{https://doi.org/10.1021/jz300456h}.

\bibitem[{\citenamefont{M{\"u}ller et~al.}(2004)\citenamefont{M{\"u}ller,
  Lupton, Rogach, Feldmann, Talapin, and Weller}}]{Mulleretal2004}
\bibinfo{author}{\bibfnamefont{J.}~\bibnamefont{M{\"u}ller}},
  \bibinfo{author}{\bibfnamefont{J.~M.} \bibnamefont{Lupton}},
  \bibinfo{author}{\bibfnamefont{A.~L.} \bibnamefont{Rogach}},
  \bibinfo{author}{\bibfnamefont{J.}~\bibnamefont{Feldmann}},
  \bibinfo{author}{\bibfnamefont{D.~V.} \bibnamefont{Talapin}},
  \bibnamefont{and} \bibinfo{author}{\bibfnamefont{H.}~\bibnamefont{Weller}},
  \bibinfo{journal}{Physical Review Letters} \textbf{\bibinfo{volume}{93}},
  \bibinfo{pages}{167402} (\bibinfo{year}{2004}),
  \urlprefix\url{https://link.aps.org/doi/10.1103/PhysRevLett.93.167402}.

\bibitem[{\citenamefont{Empedocles and
  Bawendi}(1997)}]{EmpedoclesandBawendi1997}
\bibinfo{author}{\bibfnamefont{S.~A.} \bibnamefont{Empedocles}}
  \bibnamefont{and} \bibinfo{author}{\bibfnamefont{M.~G.}
  \bibnamefont{Bawendi}}, \bibinfo{journal}{Science}
  \textbf{\bibinfo{volume}{278}}, \bibinfo{pages}{2114} (\bibinfo{year}{1997}),
  \urlprefix\url{http://science.sciencemag.org/content/278/5346/2114.abstract}.

\bibitem[{\citenamefont{Lin et~al.}(2017)\citenamefont{Lin, Dai, Pu, Deng, Niu,
  Tong, Fang, Jin, and Peng}}]{Linetal2017}
\bibinfo{author}{\bibfnamefont{X.}~\bibnamefont{Lin}},
  \bibinfo{author}{\bibfnamefont{X.}~\bibnamefont{Dai}},
  \bibinfo{author}{\bibfnamefont{C.}~\bibnamefont{Pu}},
  \bibinfo{author}{\bibfnamefont{Y.}~\bibnamefont{Deng}},
  \bibinfo{author}{\bibfnamefont{Y.}~\bibnamefont{Niu}},
  \bibinfo{author}{\bibfnamefont{L.}~\bibnamefont{Tong}},
  \bibinfo{author}{\bibfnamefont{W.}~\bibnamefont{Fang}},
  \bibinfo{author}{\bibfnamefont{Y.}~\bibnamefont{Jin}}, \bibnamefont{and}
  \bibinfo{author}{\bibfnamefont{X.}~\bibnamefont{Peng}},
  \bibinfo{journal}{Nature Communications} \textbf{\bibinfo{volume}{8}},
  \bibinfo{pages}{1132} (\bibinfo{year}{2017}), ISSN \bibinfo{issn}{2041-1723},
  \urlprefix\url{https://doi.org/10.1038/s41467-017-01379-6}.

\bibitem[{\citenamefont{Chandrasekaran
  et~al.}(2017)\citenamefont{Chandrasekaran, Tessier, Dupont, Geiregat, Hens,
  and Brainis}}]{Chandrasekaranetal2017}
\bibinfo{author}{\bibfnamefont{V.}~\bibnamefont{Chandrasekaran}},
  \bibinfo{author}{\bibfnamefont{M.~D.} \bibnamefont{Tessier}},
  \bibinfo{author}{\bibfnamefont{D.}~\bibnamefont{Dupont}},
  \bibinfo{author}{\bibfnamefont{P.}~\bibnamefont{Geiregat}},
  \bibinfo{author}{\bibfnamefont{Z.}~\bibnamefont{Hens}}, \bibnamefont{and}
  \bibinfo{author}{\bibfnamefont{E.}~\bibnamefont{Brainis}},
  \bibinfo{journal}{Nano Letters} \textbf{\bibinfo{volume}{17}},
  \bibinfo{pages}{6104} (\bibinfo{year}{2017}), ISSN \bibinfo{issn}{1530-6984},
  \urlprefix\url{https://doi.org/10.1021/acs.nanolett.7b02634}.

\bibitem[{\citenamefont{Pryor and Flatt{\'e}}(2006)}]{PryorandFlatte2006}
\bibinfo{author}{\bibfnamefont{C.~E.} \bibnamefont{Pryor}} \bibnamefont{and}
  \bibinfo{author}{\bibfnamefont{M.~E.} \bibnamefont{Flatt{\'e}}},
  \bibinfo{journal}{Applied Physics Letters} \textbf{\bibinfo{volume}{88}},
  \bibinfo{pages}{233108} (\bibinfo{year}{2006}), ISSN
  \bibinfo{issn}{0003-6951}, \urlprefix\url{https://doi.org/10.1063/1.2206679}.

\bibitem[{\citenamefont{Engel et~al.}(2001)\citenamefont{Engel, Recher, and
  Loss}}]{Engeletal2001}
\bibinfo{author}{\bibfnamefont{H.-A.} \bibnamefont{Engel}},
  \bibinfo{author}{\bibfnamefont{P.}~\bibnamefont{Recher}}, \bibnamefont{and}
  \bibinfo{author}{\bibfnamefont{D.}~\bibnamefont{Loss}},
  \bibinfo{journal}{Solid State Communications} \textbf{\bibinfo{volume}{119}},
  \bibinfo{pages}{229} (\bibinfo{year}{2001}), ISSN \bibinfo{issn}{0038-1098},
  \urlprefix\url{http://www.sciencedirect.com/science/article/pii/S0038109801001107}.

\bibitem[{\citenamefont{Liu et~al.}(2013)\citenamefont{Liu, Cao, Huang, Li,
  Guo, Laine, Breuer, and Piilo}}]{Liuetal2013}
\bibinfo{author}{\bibfnamefont{B.-H.} \bibnamefont{Liu}},
  \bibinfo{author}{\bibfnamefont{D.-Y.} \bibnamefont{Cao}},
  \bibinfo{author}{\bibfnamefont{Y.-F.} \bibnamefont{Huang}},
  \bibinfo{author}{\bibfnamefont{C.-F.} \bibnamefont{Li}},
  \bibinfo{author}{\bibfnamefont{G.-C.} \bibnamefont{Guo}},
  \bibinfo{author}{\bibfnamefont{E.-M.} \bibnamefont{Laine}},
  \bibinfo{author}{\bibfnamefont{H.-P.} \bibnamefont{Breuer}},
  \bibnamefont{and} \bibinfo{author}{\bibfnamefont{J.}~\bibnamefont{Piilo}},
  \bibinfo{journal}{Scientific Reports} \textbf{\bibinfo{volume}{3}},
  \bibinfo{pages}{1781 EP } (\bibinfo{year}{2013}), \bibinfo{note}{article},
  \urlprefix\url{http://dx.doi.org/10.1038/srep01781}.

\bibitem[{\citenamefont{Laine et~al.}(2014)\citenamefont{Laine, Breuer, and
  Piilo}}]{Laineetal2014}
\bibinfo{author}{\bibfnamefont{E.-M.} \bibnamefont{Laine}},
  \bibinfo{author}{\bibfnamefont{H.-P.} \bibnamefont{Breuer}},
  \bibnamefont{and} \bibinfo{author}{\bibfnamefont{J.}~\bibnamefont{Piilo}},
  \bibinfo{journal}{Scientific Reports} \textbf{\bibinfo{volume}{4}},
  \bibinfo{pages}{4620 EP } (\bibinfo{year}{2014}), \bibinfo{note}{article},
  \urlprefix\url{http://dx.doi.org/10.1038/srep04620}.

\end{thebibliography}

\end{document}